\documentclass[12pt]{iopart}

\usepackage{graphicx}

\RequirePackage{color}

\begin{document}

\title[Single bubble sonoluminescence]{Modeling the dynamics of single-bubble sonoluminescence}

\author{Lucas L. Vignoli\footnote{E-mail: lucas.lvig@gmail.com}, 
Ana L. F. de Barros\footnote{E-mail: ana1barros@gmail.com}, Roberto C. A. Thom\'e\footnote{E-mail: rthome@cefet-rj.br}, A. L. M. A. Nogueira\footnote{E-mail: nogue@cbpf.br}, Ricardo C. Paschoal\footnote{E-mail: paschoal@cbpf.br}, and Hil\'ario A. Rodrigues\footnote{E-mail: harg@cefet-rj.br}}

\address{Centro Federal de Educa\c{c}\~ao Tecnol\'ogica Celso Suckow da Fonseca -- CEFET/RJ \\
Departamento de F{\'i}sica -- DEPES  \\
Av. Maracan\~a 229, 20271-110, Rio de Janeiro, RJ, Brazil}
\begin{abstract}
 Sonoluminescence (SL) is the phenomenon in which acoustic energy is (partially) transformed into light. It may occur by means of many or just one bubble of gas inside a liquid medium, giving rise to the terms multi-bubble- and single-bubble sonoluminescence (MBSL and SBSL). In the last years some models have been proposed to explain this phenomenon, but there is still no complete theory for the light emission mechanism (especially in the case of SBSL). In this work, we will not address this more complicated particular issue, but only present a simple model describing the dynamical behaviour of the sonoluminescent bubble, in the SBSL case. Using simple numerical techniques within the software Matlab, we discuss solutions considering various possibilities for some of the parameters involved: liquid compressibility, superficial tension, viscosity, and type of gas. The model may be used as an introductory study of sonoluminescence in physics courses at undergraduate or graduate levels, as well as a quite clarifying example of a physical system exhibiting large nonlinearity.

\end{abstract}

\maketitle

\section{Introduction}

Sonoluminescence (SL) is an intriguing phenomenon, which consists of light emission by small collapsing bubbles inside liquids \cite{Schanz,RMP2002}. Such bubbles are created by ultrasonic waves when the pressure of the liquid is reduced relatively to the pressure of the gas present in the medium \cite{Putterman,Yu,Dellavale}. In general, the appearance of a cavity within a liquid implies the existence of a surface (the wall) dividing the region into two parts, each one occupied by a fluid: the inner cavity consisting of a gas and/or liquid vapor, and the liquid portion outside. The bubble initially expand to a maximum volume, and then collapse. At the final stages of the collapse, the gas inside the bubble radiate light. It is observed that the light emission is enhanced when atoms of noble gas are present inside the bubble. 

In the 90s, Gaitan et al. \cite{Gaitan} obtained in laboratory the necessary conditions to create and trap a single bubble the size of a few microns, levitating in a bottle of water under the action of a strong, stationary-wave sound field, emitting periodically flashes of light in each acoustic cycle. The trapping of a single sonoluminescent bubble in the liquid, yielding what is called single-bubble sonoluminescence (SBSL), and the production of repeated cycles of expansion and contraction, excited by ultrasonic acoustic waves, allowed a more accurate study of the phenomenon. This stability of the bubble made possible more detailed studies about the duration of the flash of light and the size of the bubble.

The conversion of the energy of sound waves into flashes of light occurring in such bubbles with few microns in size is an interesting field of research, if one considers the difficulties inherent to the process, the inadequacy of some theories and models proposed, and experimental limitations. Despite the existence of a wide variety of theories and models, the phenomenon is not completely explained, and there are still many open questions. Among them, we can cite the heating mechanism of the gas inside the bubble; the process of light emission; the role of temperature of the liquid in the intensity of the emitted radiation; the reason why water is one of the ideal fluids for observing the phenomenon. Up to now, the phenomenon has been extensively studied, but its detailed mechanisms remain still unclear.

The study of SL involves topics of physics such as hydrodynamics and thermodynamics, besides electromagnetism, statistical mechanics and atomic physics if one wants to go into the emitting mechanism. The dynamics of the bubble can be described by the Navier-Stokes equation  \cite{Landau}, provided the initial state of the bubble is defined by a set of physical parameters. The energy of one photon emitted in SBSL, if compared with the energy of one atom vibrating in the sound wave which gave rise to it, typically provides a value of of order $10^{12}$, which shows the high focalization of energy in this effect \cite{Crum}. The temperature in the interior of the bubble can reache thousands of degrees Kelvin during the collapse phase \cite{McNamara}.

In this work we carry out a modeling of SBSL hydrodynamics, using the radius of the bubble as the variable of interest. There are several models that describe the time evolution of the radius of the bubble. The most common are: the Rayleigh-Plesset model, the Herring-Trilling model, and the Keller-Miksis model~\cite{RMP2002,Hilgenfeldt,Sutherland}. All of these models are derived from the Navier-Stokes equation, using different simplifying assumptions~\cite{Brennen,Hammer}.

Rayleigh-Plesset equation describes the behaviour of compressible or incompressible fluids, and can therefore be used to compare the effects of the compressibility of the liquid on the time evolution of the bubble radius. Other   physical parameters like the liquid viscosity, the properties of the gas inside the bubble, and the superficial tension of the wall can be taken into account in these models. 

In order to compare the effects of these factors, we define in this work an useful parameter, the {\it damping factor}, which is defined as the ratio between the first and second highest values of the bubble radius. This parameter is relevant, since it compares two values of the radius: one before and the other one after the light emission. A detailed study of the bubble behaviour, including the variation of the wall speed as a function of time, is also carried out.

This work is addressed mainly to (under)graduate students and teachers. The subject requires the domain of calculus and many concepts from various fields of physics at intermediate level. The work can be useful in physics courses at university level, as well as an illustrative example of numerical calculus applied to a highly nonlinear phenomenon. It may also be valuable as an introductory-level text on SBSL addressed to young researchers.

The work is the result of a research project in sonoluminescence carried out together with undergraduate students.

\section{Description of the model}

The behaviour of a nonrelativistic fluid is described by the Navier-Stokes equation, which is valid at each point of the fluid and can be written as
\begin{equation}
\rho \frac{D\vec{\rm v}}{Dt} = \rho \vec{B} - \nabla {p} + \mu \nabla^2{\vec{\rm v}} \label{1},
\end{equation}
where $\rho$ is the fluid mass density, $\vec{\rm v}$ is the velocity field inside the fluid, $\rho \vec{B}$ is the resultant of body forces (e.g., gravity) per unit volume of the fluid, $p$ is the pressure field inside the fluid and $\mu$ is the fluid viscosity. The term $\frac{D\vec{\rm v}}{Dt}$ on the left-hand side of the above equation is the material derivative of the velocity of the fluid element. The material derivative is given by the operator
\begin{equation}
\frac{D}{Dt}  = \frac{\partial}{\partial t} + \vec{\rm v} \cdot \nabla  \label{2},
\end{equation}
where the first term on the right-hand side is the time derivative with respect to a fixed reference point of space (the Euler derivative), and the second term represents the changes of the field velocity along the movement of the fluid. 
The quantity $\vec{B}$ on the right-hand side stands for the acceleration originated from the body forces acting on the
fluid element, such as gravity or electromagnetic forces, for example. The second and third terms represent respectively the hydrostatic force and the viscous force, both per unit volume. 

The form Navier-Stokes equation is written above means that we are considering the fluid as Newtonian. Now, we remember that the flow inside and around the bubble is restricted to the radial direction, in other words, the problem exhibits spherical symmetry, which in fact is valid even beyond the neighborhood of the bubble, provided  the shape of the flask is spherical, a typical experimental situation. Thus, we take into account only the expansion and contraction motion of the bubble's radius. Besides, we assume as a first approach that the compressibility of the liquid is much smaller than that of the gas inside the bubble. In this case, one derives from Eq.~(\ref{1}) the Rayleigh-Plesset equation (a dot stands for one time derivative, $\dot{R}=dR/dt$, etc):
\begin{eqnarray}
  && \rho \left (R \ddot{R} + \frac{3}{2}  \dot{R}^2 \right ) = p_{\rm gas}  - P_0 - P(t)   - {4 \mu} \frac{\dot{R}}{R} - \frac{2S}{R} , \label{3} 
\end{eqnarray}
where $R(t)$ is the bubble's radius, $p_{\rm gas}(t)$ is the variable gas pressure inside the bubble ($p_{\rm gas}(t)$ is assumed to be uniform in our model), $P_0$ is the pressure of the liquid measured at any remote point from the bubble (typically, $P_0=1$~atm), $P(t)$ is the driven acoustic pressure at the point where the bubble is placed, and $S$ is the liquid surface tension at the bubble wall.   
%
%
%
$P(t)$ is assumed to be sinusoidal and starting an expansion cycle in $t=0$, that is, 
\begin{eqnarray}
 P(t) = - P_a \sin(\omega t) , \label{3.1}
\end{eqnarray}
$P_a$ being the amplitude of the driven pressure and $\omega$ the ultrasound angular frequency in resonance with the natural oscillations of the flask, such that the driven pressure generates a stationary ultrasound wave that traps the bubble at its center, on a pressure antinode.

Often, the effects of the compressibility of a liquid can be neglected in many problems of hydrodynamics. However, in the case of SBSL this approach is no longer justified, because a large amount of the acoustic energy driven to the bubble is emitted back from it to the liquid, in the form of a spherical shock wave (only a small amount is in fact converted into light!), which obviously could not exist in an incompressible medium (the acoustical wave emitted by the bubble is experimentally important, since its detection by a hydrophone signals the presence of the trapped bubble at the center of the spherical flask). It follows that when the compressibility of the liquid is considered, a new term is added to the right-hand side of Eq.~(\ref{3}), leading to the modified Rayleigh-Plesset equation  \cite{RMP2002}
\begin{eqnarray} 
  && \rho \left (R \ddot{R} + \frac{3}{2}  \dot{R}^2 \right ) =  p_{\rm gas}(t) - P_0 - P(t) 
     - {4 \mu} \frac{\dot{R}}{R} - \frac{2S}{R} +  \frac{R}{c} \frac{d}{dt} p_{\rm gas},  \label{4}
\end{eqnarray} 
where $c$ is the speed of sound in the liquid (which henceforth we will assume to be water). 

We adopt a van der Walls equation of state for describing the gas pressure inside the bubble, which reads
\begin{eqnarray}
 p_{\rm gas}(t) = \left (P_0 +  \frac{2S}{R_0} \right )  \left (  \frac{R^3_0-h^3}{R^3(t)-h^3} \right )^\gamma , \label{3.2}
\end{eqnarray}  
where $R_0$ is the static bubble radius, that is, the ambient bubble radius when it is not acoustically forced, $h$ is the characteristic van der Waals hard-core radius of the gas inside the bubble and $\gamma$ is the ratio between the specific heats of the gas at constant pressure and at constant volume (the adiabatic index). So, the gas pressure varies with time only by means of the bubble radius, $R(t)$.  It was assumed in Eq.~(\ref{3.2}) that the gas undergoes a so fast cycle of expansion and collapse that it is adiabatic. However, a more accurate analysis~\cite{RMP2002} allows one to conclude that the expansion is approximately isotermic ($\gamma\approx 1$) and only the final part of the collapse is indeed adiabatic. Our simplified model here consider the whole cycle as adiabatic.

In Eq.~(\ref{4}), the time derivative of the gas pressure is explicitly given by
\begin{equation}
\frac{d}{dt} p_{\rm gas} = -3 \gamma p_{\rm gas} \frac{R^2}{R^3 - h^3} \dot{R}.
\end{equation} 

Equations (\ref{3}) and (\ref{4}) are second-order differential equations for the radius $R(t)$, which have no analytical solution. So, in order to solve numerically these equations for a given set of system parameters, we used the mathematical program MATLAB (the corresponding program codes can be requested by email to any of the authors).

\section{Numerical solutions}

\begin{figure}[htpb]
\centering 
\includegraphics[width=.35\textheight]{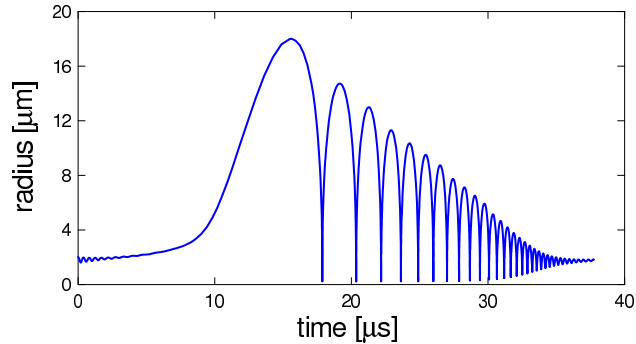}
\vspace*{-.1cm} 
\caption{(Color online) - Radius as a function of time for one (acoustically forced) cycle of the motion of the bubble. The liquid is treated as  incompressible and the bubble dynamics is described by the Rayleigh-Plesset equation (\ref{3}). \label{fig:100}}
\end{figure}

Let us first consider the case in which the compressibility of the liquid is neglected. Figure~\ref{fig:100} shows the solution of Eq.~(\ref{3}) for a given set of parameters describing some properties of the gas and the liquid, and at forcing pressure $P_a = 1.42$ atm. The parameters used are shown in Table \ref{table1}, where the noble gas argon is considered because it is well-known in SBSL literature~\cite{RMP2002} that, after some cycles, it is the only remaining constituent of the initial air bubble, due to chemical reactions that the other components participate, which carry them to the water outside. The solution shown in Fig.~\ref{fig:100} is qualitatively similar to the experimental results~\cite{Crum, RMP2002}, but quantitatively very different: the experimental afterbounces are much smaller than those that appear in Fig.~\ref{fig:100}. This is in agreement with the fact that indeed a considerable amount of energy is lost in the first contraction (collapse), which, as we have already pointed out, is due to sound radiation by the bubble, an effect that is impossible theoretically under the assumptions that led to Eq.~(\ref{3}).

\begin{table}[htbp]
\begin{center}
\caption{\label{table1} Parameters used in the numerical simulation.}
\vspace*{.3cm}
\begin{tabular}{l|l}
\hline
Superficial tension & $S=72.8 \times 10^{-3}$ N$\cdot$m$^{-1}$ \\ 
Density (water) & $\rho =$ 1000 kg$\cdot$m$^{-3}$ \\
Adiabatic index (argon) & $\gamma = 5/3$ \\
Speed of sound (water) & $c = 1,500 $ m$\cdot$s$^{-1}$ \\ 
Viscosity (water) & $\mu =1.002 \times 10^{-3}$ Pa$\cdot$s \\
Ambient pressure & $P_0 = 1.00 $ atm \\
Static radius & $R_0 = 2.0 \times 10^{-6}$ m \\
Hard core (argon) & $h = R_0/8.86$ \\
Ultrasound frequency & $\omega = 2\pi f $; $f = 1/T = 26,5$ Hz \\
\hline
\end{tabular}
\end{center}
\end{table}

\begin{figure}[htpb]
\centering 
\includegraphics[width=.35\textheight]{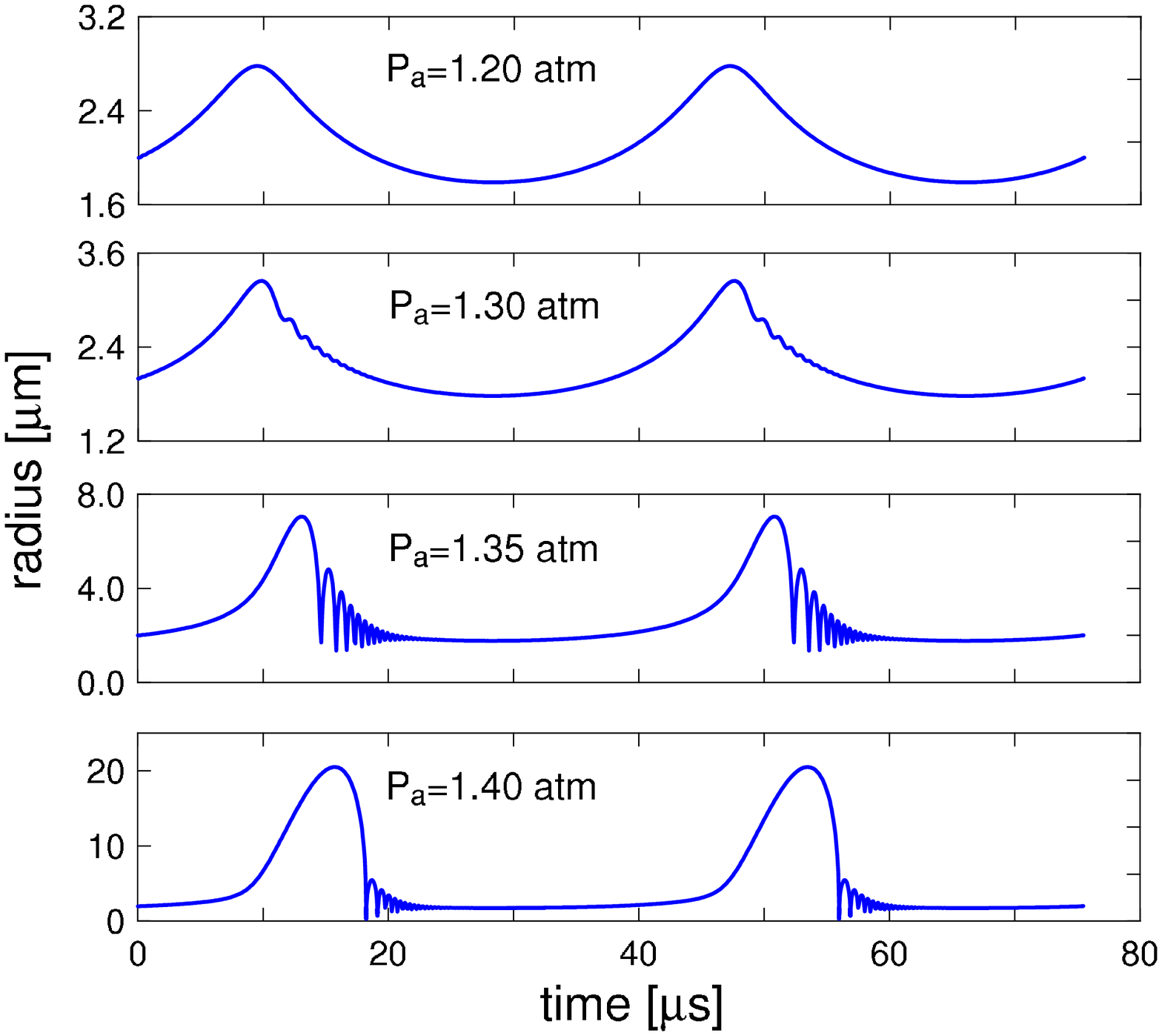}
\vspace*{-.1cm} 
\caption{(Color online) - Radius as a function of time for two (acoustically forced) cycles of the subsonic motion of the bubble. The liquid is treated as  compressible and the bubble dynamics is described by the modified Rayleigh-Plesset equation (\ref{4}). \label{fig:200}}
\end{figure}

\begin{figure}[htpb]
\centering 
\includegraphics[width=.35\textheight]{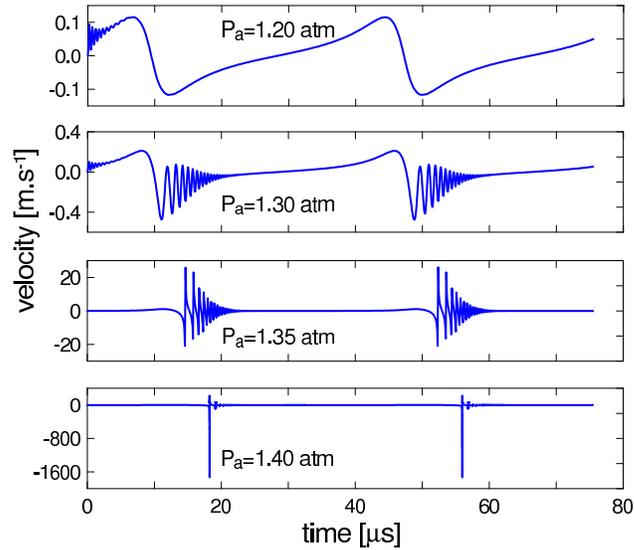}
\vspace*{-.1cm} 
\caption{(Color online) - Bubble wall velocity as a function of time for two (acoustically forced) subsonic cycles. The liquid is treated as incompressible and the bubble dynamics is described by the modified Rayleigh-Plesset equation (\ref{4}). \label{fig:300}}
\end{figure}

Figure \ref{fig:200} shows the solutions to the modified Rayleigh-
Plesset equation (20) for four different forcing pressures, and the parameters shown in Table~\ref{table1}. Under the action of the 
external forcing, the bubble radius oscillates almost sinusoidally around the equilibrium value $R_0$,   
with small amplitudes and with a period close to that of the external forcing. However, for a critical forcing pressure around $P_a = 1.35$ atm, one observes a nonlinear behaviour in the bubble dynamics.  As shown in Figure \ref{fig:300}, in this regime the motion of the bubble is subsonic, except at the forcing pressure $P_a = 1.4$ atm.  
 
\begin{figure}[htpb]
\centering 
\includegraphics[width=.35\textheight]{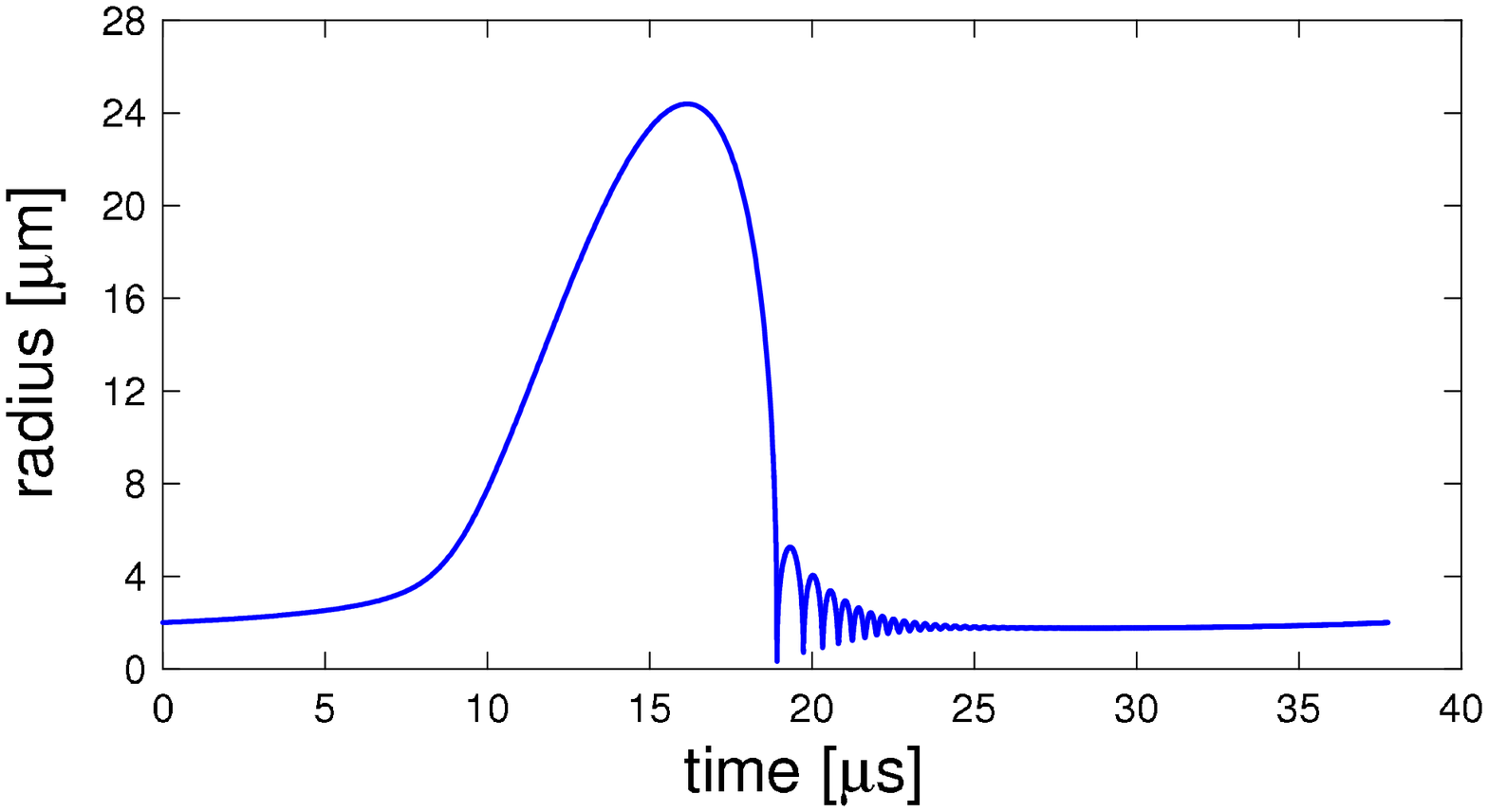}
\vspace*{-.1cm} 
\caption{(Color online) - Radius as a function of time for one (acoustically forced) cycle of the supersonic bubble collapse. The forcing pressure is $P_a=1.42$ atm. The other used model parameters are shown in Table \ref{table1}. \label{fig:400}}
\end{figure}

The sonoluminescence phenomenon is expected to occur beyond this threshold. In Fig. \ref{fig:400} we depict the solution of Eq.~(\ref{4}) for the bubble radius at the forcing pressure $P_a = 1.42$ atm, just beyond the threshold value. In this example, the initial radius is $R_0 = 2.0$ $\mu$m, which is a typical value for the radius corresponding to the mechanical equilibrium bubble under the action of an external ambient pressure (normally $ \approx $ 1 atm). In the first stage the negative forcing pressure favours a quick expansion of the bubble. This first stage ends when the the forcing pressure becomes positive. When the total pressure acting on the bubble increases, however, the bubble continues to expand, due to its own inertia, reaching the maximum radius $R \approx 10$ $R_0$ when the total pressure returns to approximately the previous value of 1 atm. At this point, the volume of the bubble is about $10^3$ of its initial value, and therefore the gas pressure is reduced by a factor of around $10^3$.  The atoms of the rarefied gas close to the wall can no more balance the external pressure of $\approx 1$ atm outside, and then the bubble collapses. This kind of calculation was firstly carried out by Lord Rayleigh in 1917~\cite{Hilgenfeldt}.

\begin{figure}[htpb]
\centering 
\includegraphics[width=.35\textheight]{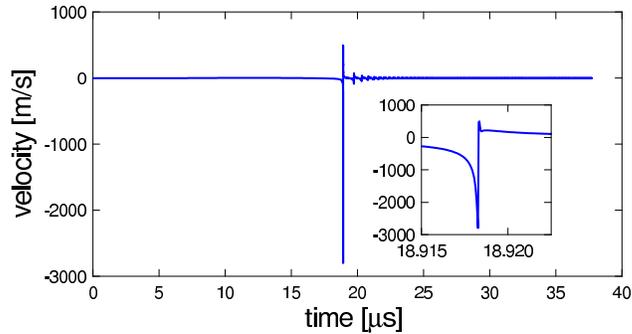}
\vspace*{-.1cm} 
\caption{(Color online) - Velocity of the bubble wall corresponding to the solution showed in figure \ref{fig:400}. \label{fig:500}}
\end{figure}

Comparing the Figures \ref{fig:100} and \ref{fig:400}, one observes that when the compressibility of the liquid is taken into account, the experimental results are better reproduced by the model \cite{Crum, RMP2002}.  In the experiments carried out in laboratory with a single bubble, it is observed that the bubble pulses synchronously with the acoustic field, expanding during a part of acoustic rarefaction cycle and collapsing during the compression phase.

A very important aspect of sonoluminescence regards the velocity of the bubble wall, especially during the collapse. At this stage, if the speed of the bubble wall is comparable or even higher than the speed of sound in the liquid and/or the gas (and this really happens in SBSL experiments), then a shock wave of relatively high intensity is generated inside each of these media. With regard to the liquid, this has been considered above, but with respect to the shock wave inside the gas, this is an issue that we will not address here, due to its complication. However, it is to be noticed that some early models for the mechanism of light emission in SBSL had this shock wave inside the bubble as a key ingredient, but nowadays these models are no more believed to be correct~\cite{RMP2002}.

The velocity of the bubble wall depends on some important characteristics, such as the properties of the gas inside the bubble, the initial bubble radius, and the properties of the liquid outside the bubble. Figure \ref{fig:500} shows the behaviour of the bubble wall speed as a function of time. The end of collapse occurs when the bubble radius approaches the van der Waals hard core of the gas. The gas reaches very high values of density (near to solid state ones!) and temperature, and then the flash of light is emitted.

\section{The effects of different parameters}

As mentioned in the Introduction, the damping factor is defined as the ratio between the first and second highest values reached by the bubble radius during the time evolution. So, the greater the losses due to dissipation in the dynamics of the bubble, the greater the damping factor. In order to investigate the effects of the physical properties of the specific gas inside the bubble (which affects the van der Waals' hard core) on the the damping factor, as well as the role of the viscosity and the superficial tension of the liquid, we have carried out several simulations. First, we consider separately the effects of each of the parameters (three simulations). Then, systematically we added the effects of the remaining parameters on the solutions of the equation of motion, providing four more simulations. The results obtained are schematically presented in Table~\ref{table2} and we invite the interested reader to send us an email in order to have access to the MATLAB codes of the program we have used.

\begin{table}[ht]
\begin{center}
\caption{\label{table2} Parameters used in the numerical simulation.}
\vspace*{.3cm}
\begin{tabular}{l|c}
\hline
{\bf Properties} & {\bf Damping factor}\\
\hline
van der Waals hard core radius, superficial tension and viscosity & 3.602 \\ 
van der Waals hard core radius & 11.434 \\
Viscosity & 11.629 \\
Superficial tension  & 4.063 \\
van der Waals hard core radius and viscosity & 11.778 \\
van der Waals hard core radius and superficial tension & 4.097 \\
Viscosity and superficial tension & 3.653 \\
\hline
\end{tabular}
\end{center}
\end{table}

We can see from Table~\ref{table2} that the superficial tension strongly affects the dynamics of the bubble, when compared with the other two properties considered here, since it substantially reduces the damping factor. According to the experimental results existing in the literature for SBSL~\cite{RMP2002}, the damping factor lies between the two extreme values shown in the table, and closer to the lower values, thus pointing out the importance of including the surface tension in the calculations.

\section{Conclusions}

In this work we have presented a model describing a single bubble sonoluminescence. We obtained solutions for the equations of the model in the cases of compressible and incompressible liquids. Comparing Figs.~\ref{fig:100} and \ref{fig:200} with the experimental results, we show that the modified Rayleigh-Plesset equation valid for a compressible liquid, given by Eq.~(\ref{4}), describes the bubble dynamics with a better accuracy~\cite{Crum}, which is due to the experimentally observed acoustic radiation from the bubble, a fact that demands the liquid compressibility to be different from zero. 

We introduced the damping factor in order to compare the effects of some  properties of the system on the dissipation of energy of the  bubble after the collapse (first compression). We pointed out that the superficial tension of the liquid can not be neglected in any model describing SBSL dynamics, as we summarized in Table~\ref{table2}.

It is worth to mention that chemical effects was not considered in the present work. The chemical factors clearly affect the  thickness of the bubble wall, thereby also contributing to the generation of a shock wave. Of course, a more realistic treatment must include such kind of ingredient. Also, as mentioned above, the process that the bubble gas undergoes is not adiabatic during the whole cycle, as we considered here, but instead isotermic during the expansion, which affects the value of the exponent $\gamma$. However, as pointed out in Ref.~\cite{RMP2002}, just modifying the numerical calculation in order to interpolate between $\gamma=1$ and $\gamma=5/3$ is not sufficient to completely describe the behaviour of $R(t)$ in all of its details.

MATLAB routines that generates our results can be requested by email to any of the authors and we hope that the simulations presented here will be valuable for teachers, (under)graduate students and young researchers in order to have a first contact with single-bubble sonoluminescence and some of its remarkable features.

\section{Acknowledgment}
The authors thank the Brazilian foundations CNPq, FAPERJ and CAPES for the financial support.

\end{document}